**Topological Quantum Statistical Mechanics and Topological Quantum Field Theories**

Zhidong Zhang

Shenyang National Laboratory for Materials Science, Institute of Metal Research, Chinese Academy of Sciences, 72 Wenhua Road, Shenyang, 110016, P.R. China

**Abstract**

The Ising model describes a many-body interacting spin (or particle) system, which can be utilized to imitate the fundamental forces of nature. Although it is the simplest many-body interacting system of spins (or particles) with $Z_2$ symmetry, the phenomena revealed in Ising systems may afford us lessons for other types of interactions in nature. In this work, we first focus on the mathematical structure of the three-dimensional (3D) Ising model. In the Clifford algebraic representation, many internal factors exist in the transfer matrices of the 3D Ising model, which are ascribed to the topology of the 3D space and the many-body interactions of spins. They result in the nonlocality, the nontrivial topological structure, as well as the long-range entanglement between spins in the 3D Ising model. We review briefly the exact solution of the ferromagnetic 3D Ising model at the zero magnetic field, which was derived in our previous work. Then, the framework of topological quantum statistical mechanics is established, with respect to the mathematical aspects (topology, algebra, and geometry) and physical features (the contribution of topology to physics, Jordan–von Neumann–Wigner framework, time average, ensemble average, and quantum mechanical average). This is accomplished by generalizations of our findings and observations in the 3D Ising models. Finally, the results are generalized to topological quantum field theories, in consideration of relationships between quantum statistical mechanics and quantum field theories. It is found that these theories must be set up within the Jordan–von Neumann–Wigner framework, and the ergodic hypothesis is violated at the finite temperature. It is necessary to account the time average of the ensemble average and the quantum mechanical average in the topological quantum statistical mechanics and to introduce the parameter space of complex time (and complex temperature) in the topological quantum field theories. We find that a topological phase transition occurs near the infinite temperature (or the zero temperature) in models in the topological quantum statistical mechanics and the topological quantum field theories, which visualizes a symmetrical breaking of time inverse symmetry.

The corresponding author: Z.D. Zhang, e-mail address: zdzhang@imr.ac.cn

# 1. Introduction

It is known that four different fundamental forces, i.e., electromagnetic, weak, and strong forces (or interactions), and gravity, naturally exist in nature. Fully understanding the behavior of these interactions has been a long-term challenge in physics and mathematics. The electromagnetic (or magnetic) interactions can be studied well within the framework of electromagnetic theory and/or quantum electrodynamics [1,2]. The electromagnetic, weak, and strong forces can be treated by the standard model as a unified theory [2–6], but the study of gravity should involve general relativity [7]. The contradiction in concepts and representations of quantum mechanisms (QM) and general relativity makes the construction of a grand unified theory for understanding the four fundamental forces extremely difficult. Another difficulty that hinders our understanding of the nature of the fundamental forces is the lack of exact solutions of many-body interacting (force) systems. Among the physical models describing the many-body interacting spin systems, the Ising model is one of the simplest models [8–10]. Therefore, the exact solution of the three-dimensional (3D) Ising model is very important for understanding other many-body interacting spin systems. Two conjectures were proposed by the author in [10] for solving the exact solution of the ferromagnetic 3D Ising model at the zero external magnetic field. Zhang, Suzuki, and March proved four theorems (Trace Invariance Theorem, Linearization Theorem, Local Transformation Theorem, Commutation Theorem) in [11], which rigorously proved Zhang's two conjectures. Furthermore, Suzuki and Zhang [12,13] rigorously proved the two conjectures by the method of the Riemann–Hilbert problem, the monoidal transformation, and the Gauss–Bonnet–Chern formula. Recently, the lower bound of the computational complexity of the spin-glass 3D Ising model was determined [14]. More recently, the exact solution of a ferromagnetic/antiferromagnetic 2D Ising model with a transverse magnetic field was derived by equivalence between the transverse-field 2D Ising model and the ferromagnetic/antiferromagnetic 3D Ising model [15]. During these procedures, we found that the 3D Ising models serve as a good platform for a sensible interplay between physical properties of interacting many-body systems, algebra, topology, and geometry. Moreover, we uncovered several characters of the 3D Ising models, which help us to develop topological quantum statistical mechanics (TQSM), in which one needs to consider contributions of nontrivial topological structures to physical properties for quantum statistical mechanics (QSM). Analogously, in some models of quantum field theory (QFT), nontrivial topological structures exist also, which can be defined as topological quantum field theory (TQFT). Although the 3D Ising model is one of the simplest many-body interacting models, the findings in our previous work [10–13] may be applicable for other many-body interacting systems. It will be extremely important to catalog the models in the same class and/or with the same mathematical structure. The aim of this work is to establish the framework of TQSM, and to set up the connection between TQSM and TQFT.

The exact solution of the ferromagnetic 3D Ising model stands up as a hard problem in physics for almost 100 years [8], since it is extremely difficult to overcome several key obstacles, such as nonlocal behavior (nontrivial topological structures), non-

Gaussian behavior (nonlinear terms of Γ-matrices), noncommutative behavior of Γ-operators [11]. In the procedures developed for solving explicitly the solution [10–13], we performed a gauge transformation or a monoidal transformation in a higher dimensional space-time to deal with the nonlocal problem, processed a linearization procedure to remove the non-Gaussian obstruction, and introduced the Jordan–von Neumann–Wigner (JNW) framework to eliminate the noncommutative hindrance. Because the 3D Ising model is a paradigm in TQSM, other models in TQSM may also possess these basic characters so that the same (nonlocal, non-Gaussian, and noncommutative) obstacles do exist for exactly solving them. Thus, we have to set up the same mathematical framework for other TQSM models as we have established for the 3D Ising model with respect to aspects of topology, algebra, and geometry. Furthermore, in certain conditions, a model in QSM can be mapped to a model in QFT so that the mathematical basis established for TQSM can be generalized to be appropriate for TQFT. This is a fact that, except for gravity, the fundamental models for the other three (i.e., electromagnetic, weak, and strong) interactions can be formulated in the same/similar forms, but with different symmetries for spins (or particles). We hope that we have ensured the validity of the concept that significant progress in solving the problems of the physics of fundamental interactions can be achieved on the basis of the 3D Ising model.

In this paper, we first review the progresses in the 3D Ising models, especially emphasizing the important roles of Clifford algebra. The results obtained for the exact solution of the ferromagnetic 3D Ising model are briefly summarized in Section 2. In Section 3, we set up the framework of TQSM, with respect to the mathematical aspects

(topology, algebra, and geometry) and physical features (the contribution of topology to physics, JNW framework, time average, ensemble average, and quantum mechanical average), by a generalization of our findings and observations in the 3D Ising models. We also focus our attention on a topological phase transition with a symmetrical breaking of time inverse symmetry near the infinite temperature, accompanied with emerging of massless gauge bosons. In Section 4, the results obtained in Sections 2 and 3 are generalized to TQFTs, in consideration of relationships between QSM and QFT. It is found that it is important to take into account the time average of the ensemble average and the quantum mechanical average of any physical quantity f in the TQSM (see Section 3), and it is important to account for the real-time average of the temperature (or imaginary time) average of a function $\varphi(x(t, \tau))$ in the TQFTs (see Section 4). It is proven that the ergodic hypothesis is violated at the finite temperature in these theories. Our work indicates clearly that it is necessary to introduce an emergent time axis in the 3D Ising model and other models in the TQSM, and to integrate in the parameter space of complex time (or complex temperature) in the TQFTs. We also find that, in the TQFTs, a topological phase transition occurs with a symmetrical breaking of time inverse symmetry near the infinite temperature, which leads to the change in the parameter space from imaginary time (real temperature) to complex time (or complex temperature)

## 2. Exact Solution of the Ferromagnetic 3D Ising Model

The Hamiltonian of the 3D Ising model is written as [8–10]:

$$H = -\sum_{<i,j>} J_{ij} S_i S_j \qquad (1)$$

Here, spins with S = 1/2 are arranged on a 3D lattice, with its lattice size N = *lmn*, where *l*, *m*, *n* denote the number of lattice points along three crystallographic directions. For simplicity, we consider only the nearest neighboring interactions $J_{ij}$. For a ferromagnetic 3D Ising model, all the interactions $J_{ij}$ are positive, and are set to be $J$, $J'$, and $J''$, respectively, for interactions along three crystallographic directions in a simple orthorhombic lattice. For a spin-glass 3D Ising model [14], the interactions $J_{ij}$ with different signs are randomly distributed, which can be set to be different, $J_{ij}$, $J'_{ij}$, and $J''_{ij}$, along the three crystallographic directions.

In this section, we focus on the ferromagnetic 3D Ising model. We introduce the generators of Clifford algebra for the 3D Ising model:

$$\Gamma_{2k-1} = C \otimes C \otimes ...... \otimes C \otimes s' \otimes 1 \otimes ... \otimes 1 \quad (k-1 \text{ times } C) \tag{2}$$

$$\Gamma_{2k} = C \otimes C \otimes ...... \otimes C \otimes (-is'') \otimes 1 \otimes ... \otimes 1 \quad (k-1 \text{ times } C) \tag{2}$$

Following the notation, due to Onsager–Kaufman–Zhang [8–11,16,17], we choose the notation: $s'' = \begin{bmatrix} 0 & -1 \\ 1 & 0 \end{bmatrix}$ (=i$\sigma_2$), $s' = \begin{bmatrix} 1 & 0 \\ 0 & -1 \end{bmatrix}$ (=$\sigma_3$), $C = \begin{bmatrix} 0 & 1 \\ 1 & 0 \end{bmatrix}$ (=$\sigma_1$), where $\sigma_j$ (j = 1, 2, 3) are Pauli matrices. The partition function of the ferromagnetic 3D Ising model at the zero magnetic field can be given as follows [8–11,16,17]:

$$Z = (2\sinh 2K)^{\frac{m \cdot n \cdot l}{2}} \cdot trace(V_3 V_2 V_1)^m \equiv (2\sinh 2K)^{\frac{m \cdot n \cdot l}{2}} \cdot \sum_{i=1}^{2^{n \cdot l}} \lambda_i^m \tag{3}$$

$$V_3 = \prod_{j=1}^{nl} \exp\{iK''\Gamma_{2j} \left[ \prod_{k=j+1}^{j+n-1} i\Gamma_{2k-1}\Gamma_{2k} \right] \Gamma_{2j+2n-1}\} = \prod_{j=1}^{nl} \exp\{iK''s'_j s'_{j+n}\}; \tag{4}$$

$$V_2 = \prod_{j=1}^{nl} \exp\{iK'\Gamma_{2j}\Gamma_{2j+1}\} = \prod_{j=1}^{nl} \exp\{iK's'_j s'_{j+1}\}; \tag{5}$$

$$V_1 = \prod_{j=1}^{nl} \exp\{iK* \cdot \Gamma_{2j-1}\Gamma_{2j}\} = \prod_{j=1}^{nl} \exp(K*C_j). \tag{6}$$

Here, variables $K = J/(k_B T)$, $K' = J'/(k_B T)$ and $K'' = J'/(k_B T)$ are introduced, instead of $J, J'$ and $J''$. K* is defined by $e^{-2K} \equiv \tanh K*$ [8–11,16–19]. We define the matrices $C_j$ and $s'_j$ as follows: $C_j = I \otimes I \otimes ... \otimes I \otimes C \otimes I \otimes ... \otimes I$ and $s'_j = I \otimes I \otimes ... \otimes I \otimes s' \otimes I \otimes ... \otimes I$. For the ferromagnetic 3D Ising model, the Clifford algebraic representation plays an important role in solving analytically its exact solution [10,11,16,17]. The difficulties arisen by the transfer matrix **V₃** (as seen from Equation (5)) for exactly solving the ferromagnetic 3D Ising model at the zero magnetic field consist of nonlocal behavior, non-Gaussian, and noncommutative operators [11]. One must take into account the contribution of nontrivial topological effect to the partition function and thermodynamic properties of the 3D Ising system. Any procedures that do not consider such contributions cannot derive a correct solution, owing to lack of important energy terms arising from topology. This is also why approximations (such as conventional low-temperature expansions, conventional high-temperature expansions, renormalization group, and Monte Carlo simulations) without consideration of nontrivial topological contributions cannot serve as a standard for adjudging the correctness of an exact solution [10–13].

**Definition 1.** *The topological structures of a ferromagnetic 3D Ising system at the zero magnetic field are constructed by the partition function Z (Equation (4)), which consists of the transfer matrices V₁, V₂, and V₃ represented in Equations (5)–(7). To illustrate the topological structures, at first, a knot γ = {Pⱼ} is constructed by the horizontal line and vertical joint line with vertex {Pⱼ} on the 3D lattice Z₃; see Figure 1, as an example. This kind of knot, which can be either nontrivial or trivial, represents the local spin alignments at the lattice points and the nearest neighboring interactions between spins (edges connecting points). The linear terms of Γ matrices in the transfer matrices V₁ and V₂ correspond to circles (or intervals), while the internal factors of nonlinear terms*

*of $\Gamma$ matrices in $V_3$ correspond to braids [12]. The circles (or intervals) and braids are attached on every lattice point. In Figure 1, we only illustrate some braids, for simplicity, not the circles (or intervals). The nontrivial topologic structures that we are interested in in this work are these braids, which represent the long-range entanglements between spins in the 3D many-body interacting spin systems [12].*

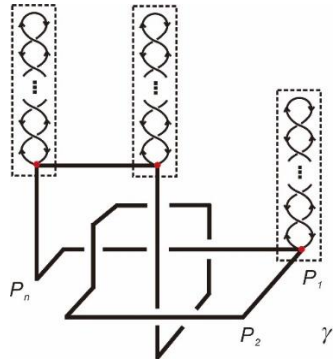

**Figure 1.** Schematic illustration of a knot $\gamma$ = {$P_j$} on the 3D lattice $Z_3$, which is connected to some braids in the transfer matrix **V₃**, while the circles (or intervals) of **V₁** and **V₂** are not shown for simplicity [12].

**Remark 1**. *The form of the knot γ is not the focus of this work, while we are interested mainly in the braids attached to the knot γ. The existence of nontrivial topologic structures in a ferromagnetic 3D Ising system at the zero magnetic field is caused by many-body interactions between spins located in a 3D lattice and the 2D character of the transfer matrices in the quantum statistical mechanism. Each factor of $i\Gamma_{2k-1}\Gamma_{2k}$ in the transfer matrices $V$ is equal to the Pauli matrix $-\sigma_k^z$, contributing a crossing topologically [12]. The crossings of the braids correspond to the $\Gamma$ matrices in internal factors of $V_3$, while the circles (or intervals) represent the linear terms in $V_1$ and $V_2$ [12]. Schematic illustration of the nontrivial topologic structures can be seen in Figure 1.*

A braid represents a long-range many-body entanglement in which all the spins in a plane are involved. Indeed, there are two kinds of contributions to physical properties of the 3D Ising model: one is from local spin alignments and another is from braids [12,13]. An equivalent representation of the nontrivial topologic structure shown in Figure 1 can be found in Figure 2.

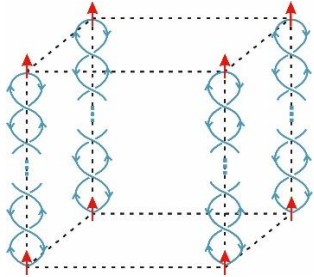

**Figure 2.** A unit cell of the 3D simple cubic Ising lattice (black dashed lines) with spins (red arrows) located at every lattice site and braids (blue curves) attached to every pair of the two nearest neighboring lattice sites along the third dimension [13]. Note that each braid corresponds to a long-range many-body entanglement between spins in a (001) plane.

The spins in Figure 2 can be mapped to crossings so that Figure 2 can be mapped to Figure 3.

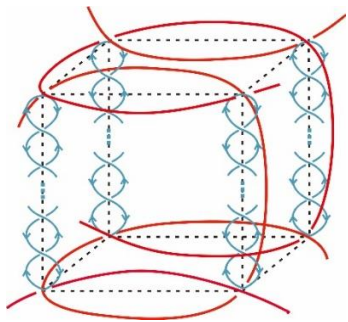

**Figure 3.** A unit cell of the 3D simple cubic Ising lattice (black dashed lines) with knots constructed from two parts, in which some crossings (red curves) are mapped from spins and some crossings (blue curves) are attached braids representing long-range many-body entanglements between spins in a (001) plane [13].

**Theorem 1**. *The nontrivial topologic problem of a ferromagnetic 3D Ising system at the zero magnetic field can be solved by introducing an additional rotation in a four-dimensional (4D) space or a (3+1)D space–time. A spin representation in $2^{n\,l\,o}$-space can be found for this additional rotation in $2n\,l\,o$-space, while the transfer matrices $V_1$, $V_2$, and $V_3$ can be represented and rearranged also in the $2n\,l\,o$-space. Namely, for the knot γ and attached braids defined in Definition 1, we have data (γ, X) which are called Knot/Clifford (K/C) data, $Z_\gamma$. For a given $Z_\gamma$, we can make a trivialization $\tilde{Z}_{\tilde{\gamma}}$.*

**Proof of Theorem 1.** Zhang, Suzuki, and March developed a Clifford algebraic approach for the ferromagnetic 3D Ising model at the zero magnetic field and rigorously proved four theorems (Trace Invariance Theorem, Linearization Theorem, Local Transformation Theorem, Commutation Theorem) [11]. By the introduction of many unit matrices to the partition function, the 3D Ising model can be expanded to be realized in a 4D space or a (3+1)D space–time. This process expands the space of spin representation from $2^{n\,l}$-space to $2^{n\,l\,o}$-space, while expanding the space of rotations from $2n\,l$-space to $2n\,l\,o$-space. Then, we can adjust the sequence of these unit matrices to isolate the (3+1)D system to be many quasi-2D subsystems. According to the property of the direct product of matrices, adding and changing positions of unit matrices can keep the trace of the transfer matrices' invariance (with compensation of a factor). Each quasi-2D subsystem with sub-transfer matrices stands for the contribution of a plane of the 3D Ising lattice and interactions with its neighboring plane. At the quasi-2D limit, we can perform a linearization process on the nonlinear terms of the transfer matrices, while splitting the space to be many subspaces. Then we can carry out a local transformation (a gauge transformation), represented by a rotation, to trivialize the nontrivial topological structures and to generalize the topological phases on eigenvectors. Clearly, the transfer matrices **V₁**, **V₂**, and **V₃** can be represented and rearranged in the 2n l o-space of rotations. Meanwhile, a spin representation in $2^{n\,l\,o}$-space can be found for this additional rotation in 2n l o-space. The Zhang–Suzuki–March approach in [11] rigorously proved Zhang's two conjectures proposed in [10] and verified the correctness of the conjectured exact solution.

Furthermore, Suzuki and Zhang [12] employed the method of the Riemann–Hilbert problem and performed the monoidal transformation to trivialize the nontrivial topological structure in the ferromagnetic 3D Ising model at the zero magnetic field. We can find a 4D manifold $\tilde{M}^4$, and K/C data $\tilde{Z}_{\tilde{\gamma}} = \{\tilde{\gamma}, \tilde{X}\}$ on $\tilde{M}$, where $\tilde{\gamma}$ is a knot and $\tilde{X} = \{\tilde{\Gamma}_j\}$, $\tilde{\Gamma}_j$ is a member of generators satisfying the following condition: we can find a trivialization mapping [12], $F: \tilde{M} \times K/C(\tilde{M}) \to Z_3 \times K/C(Z_3)$, satisfying $F(\tilde{Z}_{\tilde{\gamma}}) = Z_\gamma$. These procedures in [12] have rigorously proven Zhang's conjecture 1. □

**Theorem 2**. *Let M be a manifold, $p_0 \in M$ and let $\pi_1(M, p_0)$ be the fundamental group of M-based a t$p_0$. Consider a monodromy representation: $\rho: \pi_1(M, p_0) \to GL(M, C)$. Given the representation for the ferromagnetic 3D Ising model at the zero external magnetic field:*

$$\tilde{\rho}: \pi_1(R_g - \{a_1, ..., a_N\}) \otimes Cl(I_{3D}) \to CL^*(I_{3D})$$

*with singularities $a_i$ (i = 1,...,N). The vertex operators are introduced to represented knots for the 3D Ising model. A solution of the Riemann–Hilbert problem results in a flat connection. The three topological phases are generated on the eigenvectors of the*

*ferromagnetic 3D Ising model at the zero magnetic field, which represent the projection of the energy spectrum from 4D to 3D.*

**Proof of Theorem 2.** The Zhang–Suzuki–March approach [11] rigorously proved Zhang's two conjectures proposed in [10]. Furthermore, Zhang and Suzuki [13] applied the method of the Riemann–Hilbert problem, introduced vertex operators of knot types and a flat vector bundle, and used the Gauss–Bonnet–Chern formula to generate the topological phases on the eigenvectors of the ferromagnetic 3D Ising model at the zero magnetic field. The model with singularities of crossings of nontrivial topology can be renormalized by use of the derivation of the Gauss–Bonnet–Chern formula, in co-operation with the monoidal transforms in a 4D Riemann manifold. The ferromagnetic 3D Ising model with nontrivial topological structures can be transformed to be a trivial model on a nontrivial topological manifold. These procedures in [13] have rigorously proven Zhang's conjecture 2. □

Therefore, the partition function of the ferromagnetic 3D Ising model on a simple orthorhombic lattice at the zero magnetic field, being dealt within a (3+1)D framework, can be written as [10,16]:

$$N^{-1}\ln Z = \ln 2 +$$

$$\frac{1}{2(2\pi)^4} \int_{-\pi}^{\pi} \int_{-\pi}^{\pi} \int_{-\pi}^{\pi} \int_{-\pi}^{\pi} \ln[cosh2K cosh2(K' + K'' + K''') -$$

$$sinh2K cos\omega' - sinh2(K' + K'' +$$

$$K''')\left(\sum_{q=x,y,x} cos(\omega_q + \phi_q)\right)] d\omega' d\omega_x d\omega_y d\omega_z \qquad (8)$$

The phases $\varphi_x$, $\varphi_y$, and $\varphi_z$ are topological phases [16,20], which are generated by a local transformation, being a monodromy representation [11] as determined to be $\varphi_x = 2\pi$, $\varphi_y = \varphi_z = \pi/2$ [10,11,13]. The spontaneous magnetization M for the ferromagnetic 3D Ising model on a simple orthorhombic lattice was obtained as [10]:

$$M = \left[1 - \frac{16 x_1^2 x_2^2 x_3^2 x_4^2}{(1-x_1^2)^2 (1-x_2^2 x_3^2 x_4^2)^2}\right]^{\frac{3}{8}} \quad (7)$$

where $x_i = e^{-2K_i}$ (i = 1, 2, 3, 4), and where $K_i = J_i/(k_B T)$ (i = 1,2,3) for $K$, $K'$, and $K''$, and $K_4 = K''' = K'K''/K$. The critical temperature of the simple orthorhombic Ising lattices was given by the relation of $KK^* = KK' + KK'' + K'K''$ [10]. The critical temperature of the simple cubic Ising lattice is determined by $K^* = 3K$, i.e., $1/K_c = 4.15617384...$ The golden ratio $x_c = e^{-2K_c} = \frac{\sqrt{5}-1}{2}$ (or silver ratio $x_c = \sqrt{2}-1$) is the largest solution for the critical temperature of the ferromagnetic 3D (or 2D) Ising systems, which corresponds to the most symmetric lattice in 3D (or 2D) [10]. The critical exponents for the ferromagnetic 3D Ising model were determined to be α = 0, β = 3/8, γ = 5/4, δ = 13/3, η = 1/8, and ν = 2/3, satisfying the scaling laws [10].

It should be noted that nonlocal (topological), non-Gaussian (nonlinear), and noncommutative behaviors of Γ-operators are acting on all the terms of operators of spins with $j = 1, ..., nl$ for the 3D Ising model (see the transfer matrices Equations (5)–(7)). If one considers that the periodic boundary condition along one direction has been applied already, all the N (= lmn) spins located at the 3D lattice are involved for calculations of bulk thermodynamic properties (see the partition function Equation (4)). One must deal with all these nonlinear terms of Γ-matrices in the partition function, as well as the transfer matrices for bulk, while the effects of boundary conditions can be neglected. It is well known that, in the thermodynamic limit for the bulk free energy per site, the boundary terms have no effect, owing to the Bogoliubov inequality, as the surface-to-volume ratio vanishes for the infinite system. Therefore, to derive an exact solution in thermodynamic limit, the contributions of boundary conditions are negligible.

**Remark 2**. *For judging the correctness of an exact solution, there is a criterion that the critical point of the 3D simple cubic Ising model must be much higher than that ($1/K_c = 3.6409569...$) of the 2D triangular Ising model. The 2D triangular Ising model is equalized to the 2D square Ising model with one next-nearest neighboring interaction. Assuming that the third interaction of the simple cubic Ising model is the same as the next-nearest neighboring interaction of the latter model, the only difference between the two models is whether the third interaction lies in the 2D (001) plane along [110] direction or points to the third dimension along [001] direction in the 3D space. Clearly, there must be an additional contribution of space (topology) in the 3D Ising lattice. Therefore, the difference between the critical points of the simple cubic lattice and the triangular lattice (or the square Ising model with one next-nearest neighboring interaction) must be positive ($\Delta T_c = T_c^{cube} - T_c^{triangle} > 0$), which originates from the topological contribution of the 3D lattice. Obviously, one can exclude the solution of $K^* = 2K$, resulting in $1/K_c = 3.28203585...$ (i.e., $K_c = 0.30468893...$) for the simple cubic Ising model, since it is even lower than that ($1/K_c = 3.6409569...$) of the triangular Ising model. For detailed discussion, readers refer to page 5334 of [10].*

The 3D Ising model can be mapped to the 3D $Z_2$ gauge lattice theory [21,22], which can be used to model the interactions between Ising spins (belonging to electromagnetism, however, with polarization). Solving the 3D Ising model can help to understand some fundamental interactions of particle physics. This is because the 3D $Z_2$ gauge lattice theory can be generalized to other gauge lattice theories with different symmetries for electromagnetic interactions, weak interactions, and strong interactions. Furthermore, the action of these gauge lattice theories can be mapped to the kinetic (i.e., gauge) terms of quantum field theories with different symmetries for these interactions, with ultraviolet cutoff.

## 3. Topological Quantum Statistical Mechanics

In this section, we first define topological quantum statistical mechanics (TQSM) and describe its important characters, with regards to the mathematical aspects (topology, algebra, and geometry) and physical features (the contribution of topology to physics, JNW framework, time average, ensemble average, and quantum mechanical average). Then, we demonstrate four corollaries and a theorem for the models cataloged into TQSM.

**Definition 2**. *The topological quantum statistical mechanics (TQSM) deals with specific models in the quantum statistical mechanics (QSM), in which many-body interactions existing in spins located at a 3D lattice cause the nontrivial topological structures and the long-range spin entanglement.*

In the TQSM, we deal with statistical models with the following characters: (1) *Topology:* the nontrivial topological structures exist in the transfer matrices for the partition function, which are caused by a combination of the planar character of the transfer matrices and the 3D lattice arrangement of many-body interacting spins. (2) *Algebra:* Clifford algebra is introduced to represent the interactions between spins in the Clifford algebra space in order to illustrate the nontrivial topological structures. Jordan algebra is introduced to deal with the noncommutative relations of $\Gamma$-operators in the transfer matrices. (3) *Geometry:* the geometry phases appear in the eigenvectors of the many-body interacting system due to the parallel transport on the Riemann surface in manifold during the topological transformation. The models in QSM with the characters above can be cataloged into TQSM, which include the 3D Ising model, the 3D $Z_2$ lattice gauge theory, the Abelian U(1) lattice gauge theory, the non-Abelian SU(2) lattice gauge theory, and the non-Abelian SU(3) lattice gauge theory. The 3D $Z_2$ lattice gauge theory is dual to the 3D Ising model, while the latter three lattice gauge theories are generalizations of the 3D $Z_2$ lattice gauge theory [21,22].

According to the topology theory [23–30], it is well known that, in the models of QSM, a spin located at a lattice can be mapped to a crossing of a knot so that a one-to-one mapping exists between a knot and a spin lattice. A state of the knot diagram is very analogous to the energetic states of a physical system. The Kauffman bracket polynomial for a knot equals the partition function of a Potts spin model (including an Ising spin model), but up to an irrelevant multiplicative factor. It is noted that the Kauffman and Jones bracket polynomials in the topology coincide when $A^{1/4} = t$ [23–30]. In the models of TQSM, besides the component of spin local alignments in the partition function, there is an additional component of nonlocal and global effects of braids caused by many-body interactions and 3D dimensionality. The braids, as represented in Figures 1–3 for the 3D Ising model as an example, illustrate the long-range entanglement of spins in a plane. Caused by the planar character of the transfer matrices in QSM and the arrangement of interacting spins in a 3D lattice, it, indeed, represents a nontrivial topological structure. It should be noted that the nontrivial topological structures in some models in TQSM may be much more complicated than what we illustrated in Figures 1–3, with more complex braids due to that the spins with U(1), SU(2), or SU(3), which may result in more complicated transfer matrices. As uncovered in [13,16], a nontrivial topological basis of a physical system can be transformed to a trivial topological basis by a transformation (a rotation), and vice versa. As long as nontrivial knots or links exist in a many-body interacting spin system, a rotation matrix representing such a transformation intrinsically and spontaneously always exists, no matter how complicated the nontrivial topological structures are. Meanwhile, the transformation changes the gauge of local systems and generates the topological/geometric phases in the eigenvectors (and the partition function) of these many-body interacting spin systems [10–13].

The following corollary is validated:

**Corollary 1.** *The topological quantum statistical mechanics must take into account the contribution of nontrivial topological structures to physical properties of the many-body interacting spin system. This can be accomplished by either of the following processes:*

*(1) Performing a gauge transformation in one-dimensional higher space, changing the gauge of spins by the transformation while it generalizes the phase factor on wave functions (or fields) of the system [11].*

*(2) Performing a monoidal transformation, resulting in a trivialization mapping with a monodromy representation [12,13].*

The Jordan algebra of $A \circ B = \frac{1}{2}(AB + BA)$ [31,32] and the JNW theorem [33] provide the mathematical basis of quantum mechanics (QM), which are compatible with Schrödinger's wave question, quantization in the Hilbert space, Heisenberg's uncertainty principle, Bohr's concept of complementarity [34], Pauli exclusion principle, and Bose statistics [35]. The JNW theorem has been applied to various quantum mechanics theories [36–39]. According to Definition 2, the TQSM deals with specific models with nontrivial topological structures in the QSM, which should be set up within the framework of the QM. The logic chain can be demonstrated briefly as follows: QM ⊃ QSM ⊃ TQSM and, because JNW for QM, we have JNW for QM → JNW for QSM → JNW for TQSM. Thus, the JNW framework being the mathematical basis of quantum mechanics is the sufficient condition that it is also the mathematical basis of topological quantum statistical mechanics.

As pointed out in [40–43], the present author gave quaternion-based 3D quantum models for order–disorder transitions in simple orthorhombic Ising lattices [10], based on JNW procedure [33]. It is necessary to apply the Jordan algebra in order to overcome the difficulty of noncommutative Γ-operators/matrices in the 3D Ising model. With commutative operators/matrices of Jordan algebra, the (3+1)D Ising model in a certain order, as the evolution of the system in space–time, can be transformed into a 3D Ising model with time average and vice versa [11]. The commutativity of transfer matrices with Jordan algebra and generalized Yang–Baxter equation together ensure the integrability of the 3D Ising model [11,16]. The 3D $Z_2$ lattice gauge theory is constructed by interaction between spins around each plaquette, with $Z_2$ symmetry for Ising spins. The 3D $Z_2$ lattice gauge theory is dual to the 3D Ising model [21,22], which must be set up on the JNW framework also. The 3D U(1), SU(2), or SU(3) lattice gauge theory has the same formulation also for interaction between spins around each plaquette, except for different symmetry U(1), SU(2), or SU(3) of operators of spins. The operators of spins in the Abelian U(1) lattice gauge theory are commutative, but similar to the case that, in the 3D $Z_2$ lattice gauge theory (as well as in the 3D Ising model), the Γ-matrix in the Clifford algebra representation satisfies anticommutative relations, which must be dealt with by application of Jordan algebra. The operators of spins in the 3D non-Abelian SU(2) or SU(3) lattice gauge theory are noncommutative, and the Γ-matrices are anticommutative; thus, the Jordan algebra is also needed to build the JNW framework. Analogously, it can be mapped into a model of the many-body interacting spins with symmetry U(1), SU(2), or SU(3) in the TQSM [21,22]. As the generalizations of the 3D Ising model (and the 3D $Z_2$ lattice gauge theory), these models in TQSM must be built up also on the JNW framework with application of Jordan algebras. Thus, the JNW framework being the mathematical basis of TQSM is the necessary condition for studying the many-body interacting spin models in 3D.

We arrive at the following corollary:

**Corollary 2.** *The topological quantum statistical mechanics must be set up on the Jordan–von Neumann–Wigner framework, with application of Jordan algebras for multiplication of operators, which ensures the integrability of the system.*

In the classical statistical mechanics, Boltzmann's fundamental hypothesis is that all microstates are equally probable, and it was assumed that two ensembles (time ensemble and statistical ensemble) lead to the same results [44]. The time ensemble consists of discrete time-out of the time development of a physical system, while the statistical ensemble consists of many systems at one instant in time [44]. The time average can be formulated in the form of:

$$<f>_t = \lim_{T \to \infty} \frac{1}{2T} \int_{-T}^{+T} f(p,q,t) dt \tag{8}$$

Under the ergodic hypothesis, the relations between the averages are discussed as follows [45]: because the ensemble in statistical mechanics is a stationary one, the ensemble average of any physical quantity $f$ is independent of time; accordingly, it also equals the time average of the ensemble average of $f$. In principle, the processes of time averaging and ensemble averaging are completely independent. So, it is thought that the order for performing these averages can be reversed without causing any change. Namely, the time average of the ensemble average of $f$ is identical to the ensemble average of the time average of $f$, and also to the long-time average of $f$ that is the same value obtained through experiment. Therefore, with the ergodic hypothesis, we have the following consequence: the ensemble average of any physical quantity f is identical to the value obtained by an appropriate measurement on the given system [45].

The situation in QSM becomes much more complicated than in classical statistical mechanics [45]. One needs to introduce the concepts of QM, such as wavefunction, probability amplitude, expectation value, etc. [45], to inspect the ergodic hypothesis. In this case, an ensemble is constructed by N identical systems with N >> 1, characterized by a common Hamiltonian $\hat{H}$. The wavefunction ψ($\mathbf{r}_i$,t) characterizes the physical states at time t of the various systems in the ensemble, where $\mathbf{r}_i$ denotes the position coordinates relevant to the system, while $\psi^k(\mathbf{r}_i,t)$ (with k = 1,2, …, N) is the normalized wavefunction for the kth system of the ensemble [45]. The time variation of the wavefunction $\psi^k(t)$, which is expressed in terms of a complete set of orthonormal functions $\varphi_n$ with coefficients $a_n^k(t)$, is controlled by the Schrödinger equation. It should be noted that the number $\left|a_n^k(t)\right|^2$ determines the probability that a measurement at time t finds the kth system of the ensemble in the particular state $\varphi_n$. The density operator is described by the matrix elements [45]:

$$\rho_{mn}(t) = \frac{1}{N}\sum_{k=1}^{N}\left|a_m^k(t)a_n^{k*}(t)\right| \tag{9}$$

The matrix element $\rho_{mn}(t)$ represents the ensemble average of the quantity $a_m(t)a_n^*(t)$. The diagonal element $\rho_{nn}(t)$ denotes the ensemble average of the probability $\left|a_n(t)\right|^2$ (i.e., a quantum mechanical average). Thus, In QSM, we must encounter a double-averaging process with respect to the probabilistic aspect of the wavefunction and the statistical aspect of the ensemble. The quantity $\rho_{nn}(t)$ accounts for the probability that one finds a system, which is chosen at random from the ensemble at time t, to be in the state $\varphi_n$, and we have $\sum_n \rho_{nn}(t) = 1$ [45]. Therefore, in

QSM, we must encounter a double-averaging process consisting of ensemble average and quantum mechanical average, in consideration of the ergodic hypothesis.

However, in statistical mechanics and QSM, the ergodic hypothesis can be rigorously proven only in very few, largely artificial cases. It does not even apply in some instances [46]. Counter examples are easily constructed in one dimension, where the existence of numerous constants of the motion prevents the initial variable from becoming randomized in so-called integrable models. It has been known that, in its original form, the ergodic hypothesis cannot be strictly maintained. Meanwhile, its application, even as a working hypothesis, would oppose statistical ensemble of their full function of representing the relative probabilities for all the different kinds of states, which include the unusual ones [47]. Except for the trivial one-dimensional case, a mechanical system cannot possibly obey this ergodic hypothesis in its original strict form, because it requires the phase point to pass precisely through every point in the phase space compatible with the system's energy. In some examples, the departure from the consequence of the ergodic hypothesis can be very great. The ergodic theorem, ergodic theory, and statistical mechanics have been extensively investigated for various physical systems [48–55], and nonergodicity has been found in several spin systems [56–58]. The ergodic hypothesis cannot be proven for the 3D many-body interacting spin systems because of the existence of nonlocality and nontrivial topological structures, as in the 3D Ising model [10,11,16].

In our previous work [11], it was shown that it is necessary to introduce the time average for solving the exact solution of the ferromagnetic 3D Ising model at the zero magnetic field, which proves that the ergodic hypothesis is violated at the finite temperature in this system and it is dealt with in (3+1) dimensions. At the infinite temperature, where the interactions are negligible, one may use only the ensemble average (see detailed discussion in proof of Theorem 3, below), and the ergodic hypothesis can be held. The 3D $Z_2$ lattice gauge theory can be mapped equivalently to the 3D Ising model, and thus the ergodic hypothesis is violated also in the 3D $Z_2$ lattice gauge model. The Abelian U(1) lattice gauge theory, the non-Abelian SU(2) lattice gauge theory, and the non-Abelian SU(3) lattice gauge theory are generalizations of the 3D $Z_2$ lattice gauge model. Though they have different spin symmetries with the 3D $Z_2$ lattice gauge model, these models in the TQSM possess some characters that are the same as those of the 3D $Z_2$ lattice gauge model (and the 3D Ising model), such as noncommutative operators (of Clifford algebras), nonlocality, and nontrivial topological structures that are caused by dimensionality and many-body interactions in a 3D lattice. All of these characters in the mathematical structure require the time average with monodromy representation in (3+1) dimensions. Therefore, the ergodic hypothesis is violated at the finite temperature also in these TQSM models.

Summarizing the procedures above, we have demonstrated the following corollary:

**Corollary 3.** *The ergodic hypothesis is violated at the finite temperature in the topological quantum statistical mechanics.*

**Remark 3.** *Corollaries 2 and 3 above are complementary in a sense that the JNW framework provides the mathematical basis of the models in TQSM with application of Jordan algebra in order to overcome the difficulty of noncommutative operators, which actually performs a time average for a physical quantity, violating the ergodic hypothesis.*

According to Corollary 3 above, in TQSM systems (such as the 3D Ising model), the ergodic hypothesis is not satisfied at the finite temperature. We have to make the time ensemble, not only the statistical ensemble and quantum mechanical average for a physical quantity [10,11,16,20,59]. Indeed, when the ergodic hypothesis is violated, we must reconsider the relations between time average, ensemble average, and quantum mechanical average. In this case, the ensemble under study cannot be considered as a completely stationary one, and it can be treated as a quasi-stationary one. Thus, the ensemble average of any physical quantity f is not independent of time; accordingly, we must take the time average of the ensemble average and the quantum mechanical average of f. The order for the processes of time averaging and ensemble averaging can be reversed without causing any change, because the operators commute with the use of Jordan algebra in the JNW framework. The time average of the ensemble average and the quantum mechanical average of f can be represented as the ensemble average and the quantum mechanical average of f in terms with multiplication of Jordan algebra. Namely, the time average of the ensemble average and the quantum mechanical average of f is equal to the ensemble average and the quantum mechanical average of the time average of f. Either of these equals the long-time average of f that equals the value one expects to obtain through experiment. For TQSM, we have:

$$<\rho_{nn}(t)>_t = \left\langle \frac{1}{N}\sum_{k=1}^{N}\left|a_n^k(t)\right|^2 \right\rangle_t = \frac{1}{N}\sum_{k=1}^{N}\left|\left\langle a_n^k(t)\right\rangle_t\right|^2 \tag{10}$$

Therefore, in TQSM, the time average of the ensemble average and the quantum mechanical average of any physical quantity f is equal to the value obtained by an appropriate measurement on the given system. Thus, these lattice gauge theories, such as the 3D $Z_2$ lattice gauge model, the Abelian U(1) lattice gauge theory, the non-Abelian SU(2) lattice gauge theory, and the non-Abelian SU(3) lattice gauge theory, should follow the 3D Ising model to be dealt with in the JNW framework and by the time average of the ensemble average and the quantum mechanical average. For these lattice gauge theories an additional dimension being time is introduced to satisfy the JNW framework.

With the preliminaries above, we can claim the following corollary:

**Corollary 4.** *As the ergodic hypothesis is violated in the topological quantum statistical mechanics, the time average of the ensemble average and the quantum mechanical average of any physical quantity f equals the value one expects to obtain on making an appropriate measurement on the given system of topological quantum statistical mechanics, which is equal to the ensemble average and the quantum mechanical average of the time average of the physical quantity f.*

Next, we prove the following theorem:

**Theorem 3**. *In models in the topological quantum statistical mechanics, a topological phase transition occurs near the infinite temperature, which visualizes a symmetrical breaking of time inverse symmetry, accompanied with emerging of massless gauge bosons.*

**Proof of Theorem 3.** In our previous work [10–13], we revealed that there is a topological phase transition at/near the infinite temperature with changing temperature. During the phase transition, the topological phases on the eigenvectors are altered. The states and the phase transition at/near the infinite temperature can be described in detail as follows.

At the infinite temperature ($\beta = 1/(k_B T) = 0$, $T = \infty$), the state is that the system is most chaotic and disordered, while the interaction is neglected as, compared with the infinite temperature, no nontrivial topological structure exists. Such an extremely disordered state with trivial topological structure exists in the temperature region (T =

∞–∞⁻). This state is kept until the temperature lowers to near the infinite temperature (β = ε → 0, i.e., β = 0⁺, T = ∞⁻), where the topological phase transition occurs. At the finite temperature, the state is that the interaction is not negligible, compared with temperature, and nontrivial topological structures exist. The topological phase transition at/near the infinite temperature does not contradict with the Lee–Yang Theorem [60,61] for phase transitions: $z = \exp\left(-\frac{2H}{k_B T}\right) = 1$, where either the zero magnetic field (H = 0) or the infinite temperature (T = ∞) may satisfy z = 1. The topological phase transition near the infinite temperature shows a symmetrical breaking for the time-reversal symmetry. In the temperature region (T = ∞–∞⁻), the state of the system at every time is the same, showing the continuous symmetry with time, which is accounted for only by the ensemble average. However, at the finite temperature, the state of the system evolves with time, which must be accounted for by both the time average (from zero to infinite time) and the ensemble average. In this sense, time emerges spontaneously and naturally, with the time arrow flowing from zero to infinity. A gap exists in between the initial state and the final state of the system with time evolution, up on the monodromy representation (or the parallel transport). That is, the time-reversal symmetry is broken down due to the topological phase transition near the infinite temperature, which is consistent with the second law of thermodynamics. According to the Noether's theorem [62], a symmetry corresponds to a conservation law of a physical quantity. The breaking of the time-reversal symmetry corresponds to the breakdown of the energy conservation law with decreasing temperature, due to a topological phase transition near the infinite temperature, which may illustrate the initial states of our Universe.

According to the duality between high- and low-temperature phases [21,22], the topological phase transition near the infinite temperature (T = ∞⁻) can be mapped into a topological phase transition near zero temperature (T = $0^+$). In the temperature region (T = 0–$0^+$), the state is that all the spins are completely orderly aligned, with a trivial topological structure. At the finite (above zero) temperature, some spins become misaligned so that nontrivial topological structures emerge. Therefore, besides the quantum phase transition at zero temperature with change in the magnetic field (or pressure, the electric field, etc.), a topological phase transition occurs near zero temperature (T = $0^+$) with changing temperature. Such a topological phase transition may prohibit the system from approaching the absolute zero if the system is cooling down by topological protection. It does not contradict with the third law of thermodynamics.

The spontaneous symmetry breaking was first discovered by Nambu [63,64], which was inspired by the superconductivity theory developed by Bardeen, Cooper, and Schrieffer [65], and the apparent violations of gauge invariance were explored on the physical models. In the 3D Ising model, usually one is interested in the spontaneous symmetry breaking corresponding to the phase transition at the critical point $T_c$, where an energy gap is opened. The results obtained in this work (and our previous work [10–13]) reveal that, besides the phase transition at the critical point $T_c$, another topological phase transition occurs near the infinite temperature (or the zero temperature). According to the Nambu–Goldstone theorem [63,64,66], it will be possible that a massless bosonic particle may generate near the infinite temperature (or the zero temperature) by such a topological phase transition, and the dynamics of this particle can be described by Yang–Baxter relations [67]. Because the topological phase transition near the infinite temperature is accompanied by the acting of interactions of the systems, the Nambu–Goldstone bosons are the particles that carry the interactions. Therefore, we expect that the Nambu–Goldstone bosons emerging near the infinite temperature can be polarized photons in the 3D $Z_2$ lattice gauge theory, photons in the Abelian U(1) lattice gauge theory for electromagnetism, massless gauge bosons in the non-Abelian SU(2) lattice gauge theory (without Higgs terms [68–72]) for weak interactions, and gluons in the non-Abelian SU(3) lattice gauge theory for strong interactions. The massless gauge bosons for the non-Abelian SU(2) lattice gauge theory are generated near the infinite temperature, which is much higher than the energy for the generation of massive Higgs bosons. These observations do not contradict with the standard model [68–81]. In the low-temperature region near zero, the topological phase transition results in the generation of dislocations (the directions of some spins point oppositely with most spins) in the 3D $Z_2$ lattice gauge theory and quasi-particles (collective excitations, analogous to spin waves) in other lattice gauge theories. □

**Remark 4.** *The mathematical structures of the TQSM are realized as a result of interplay between algebra, topology, and geometry.*

This remark is evident as a subsequence of our previous work [10–13,16]. In the 3D Ising model, we have already shown that it can serve as a good platform for interplay between algebra, topology, and geometry [10–13,16]. The 3D Ising model can be described by Clifford algebra for representation of transfer matrices [10,11], Knot/Clifford algebra for representation of knot structures [12,13], Jordan algebra with the JNW framework [11], quaternion algebra for constructing quaternionic eigenvectors [10–13,16], and Lee algebra for representing rotations. The nontrivial topologic structures are found to exist in the 3D Ising model, while two kinds of components (spin alignments and braids/knots) contribute to the partition function and thermodynamic physical properties of the 3D Ising model [10–13,16]. The Riemann–Hilbert problem method is employed to solve the differential equation with singularities.

The 4D Riemann manifold is introduced to realize the knots on the Riemann surface, to formulate the representation of the Riemann-Hilbert problem, to apply the monoidal transformations at the knot intersection (singular) points, and to produce the trivialization of knots [12,13]. We need to perform a topological transformation, as well as a gauge transformation [11], or a monoidal transformation with a monodromy representation and the Gauss–Bonnet–Chern formula [12,13] to trivialize the nontrivial topological structure and to generalize the topological phases on eigenvectors. A detailed discussion about interplay between algebra, topology, and geometry can be found in [13]. Because other models in TQSM have the basic characters of the 3D Ising model but with different symmetries, they must be dealt with also by interplay between algebra, topology, and geometry, with the similar procedures we have performed for the 3D Ising model [10–13,16].

It is worth noting a recent advance [43] in the 3D Ising lattice with the Galois extension structure of the nonion algebra: Ławrynowicz et al. [43] investigated ternary and binary structures of su(3) and identified the construction of the collection of two ternaries with the collection of three binaries by extending aspects of Ising–Onsager crystal statistics [9] to three dimensions [10], by considering binary and ternary crystalline structures, and by using the Galois extension structure of the nonion algebra. They highlighted that the approach developed in [10,11,43] is applicable to quarks and elementary particles, including introduction of colors, and suggested an analysis of three quaternaries vs. four ternaries, involving duodevicenion and/or quindenion algebra.

**4. Topological Quantum Field Theories**

**Definition 3**. *The topological quantum field theories (TQFTs) are specific models in the quantum field theories (QFTs), in which the nontrivial topological structure and the long-range spin entanglement exist due to the many-body interactions of spins in spacetime.*

The quantum field theories are central topics in high-energy physics and nuclear physics, which serve as fundamental models describing interactions and processes between elementary particles (or spins). For instance, the $\lambda\varphi^4$ scalar field model was developed to study pion–pion scattering and the π–π interaction [82–89], properties of hadrons [90,91], etc. The models, which are cataloged into TQFTs, include the $\varphi^4$ scalar field theory, the Abelian gauge field theory, the Yang–Mills SU(2) gauge field theory, and the Yang–Mills SU(3) gauge field theory [2–6,92–98]. These gauge field theories with gauge invariance possess kinetic and potential terms, in which the kinetic terms are pure gauge terms and the potential terms are Higgs(-like) terms [63,64,68–72,77,78]. The kinetic terms in these gauge field theories differ in the spin symmetries. We may consider the models with pure gauge terms only or with additional Higgs terms. In TQFTs [2–6,92–98], the nontrivial topological structures exist, which are induced by the gauge terms in 3D. Thus, one has to deal with them by specific algebraic and topological approaches and the topological phases may emerge, which are similar to Berry phase [2–6,92–98]. For more information about algebra, geometric, and topological issues of TQFTs, readers refer to [95–109]. The present work shall reveal that these gauge field theories must be dealt with in the JNW framework and in the parameter space of complex time (or complex temperature). Moreover, we shall discuss the contribution of nontrivial topological structures to physical properties of these theories.

We first choose a $\varphi^4$ scalar field theory as an example, and then generate our conclusion to other TQFTs. The Lagrangian of a $\varphi^4$ scalar field theory (or Landau–Ginzburg model) in D spacetime can be expressed as [21,110]:

$$L = \int d^{(D-1)}x \left[ \frac{1}{2}\left(\frac{\partial \phi}{\partial t}\right)^2 - \frac{1}{2}(\nabla \phi)^2 - \frac{1}{2}\mu_0^2 \phi^2 - \lambda_0 \phi^4 \right] \tag{11}$$

with a formal expression for the path integral. The Lagrangian consists of kinetic terms and potential terms. These expressions can be extended to analytically continue to imaginary time. The theory can be formulated on an anisotropic spacetime lattice by replacing integrals by sums and derivatives with discrete differences to obtain the theory's Euclidean action S on the 4D lattice, the temporal integral of the Lagrangian. The path integral Z of the lattice theory is obtained from the action S, which can be thought of as the partition function of a 4D statistical mechanics problem. By the mapping above (see Equations (3.28)–(3.38) in [21]), we have a statistical mechanics problem on a symmetric lattice and field-theoretic formulation using operators $\hat{\phi}(n)$, $\hat{\pi}(n)$, and $\hat{H}_s$ defined on a symmetric lattice, which provides the theory with an ultraviolet cutoff. The partition function of the $\varphi^4$ model equals a sum over the possible configurations of the field $\varphi$ [110]:

$$Z = \int [d\phi] \exp\left\{ -\int d^{(D-1)}x \left[ \frac{1}{2}(\nabla \phi)^2 + \frac{1}{2}\mu_0^2 \phi^2 + \lambda_0 \phi^4 \right] \right\} \tag{12}$$

The partition function of a D-dimensional classical model is very much similar to the generating function of a D space–time dimensional quantum field in the Euclidian formulism (with ultraviolet cutoff [21]). Thus, both the models have the critical behaviors in the same universality class. By inspecting the resemblance between the evolution operator in QFT and the density operator in QSM, and by the mapping of β

= $(k_B T)^{-1}$ → it = τ, one can determine the temperature–time duality in the 3D Ising model [110–113] and other models in TQFT.

The kinetic terms in the (3+1)D $\varphi^4$ scalar field theory can be mapped to the exchange terms in the 3D Ising model (and the 3D $Z_2$ lattice gauge theory), which show the nontrivial topological structures in 3D, as caused by the many-body interactions in a 3D lattice [10–13]. The potential terms, namely, the Higgs-like terms, have the $Z_2$ symmetry that is the same as the symmetry of Ising spins. One may visualize that the Ising model consists of a double-well potential at each site [21], while nearest-neighbor sites are coupled together in the usual way. The Hamiltonian of Ising systems is invariant under the operation x → −x. So, the (3+1)D $\varphi^4$ scalar field theory is in the same class of the 3D Ising model (and the 3D $Z_2$ lattice gauge theory), which is also topologically nontrivial, as induced by the many-body interactions in 3D space. Analogously, in other topological quantum field theories, such as the Abelian gauge field theory, the Yang–Mills SU(2) gauge field theory, and the Yang–Mills SU(3) gauge field theory, the kinetic (i.e., pure gauge) terms have the similar characters to those of the 3D Ising model and the $Z_2$ gauge field theory. For instance, for weak coupling, the action of Abelian U(1) lattice gauge theory (Equation (6.2) in [21]) can be reduced to the Euclidean action of electromagnetism (Equation (6.15) in [21]): $S \approx \frac{1}{4}\int d^4x F_{\mu\nu}F_{\mu\nu}$. For classical smooth fields, the action of non-Abelian SU(2) lattice gauge theory (Equation (8.3) in [21]) can be reduced to the classical Euclidean action of pure Yang–Mills SU(2) fields $S = \frac{1}{4}\int d^4x (F^i_{\mu\nu})^2$ with $F^i_{\mu\nu} = \partial_\mu A^i_\nu - \partial_\nu A^i_\mu - g\varepsilon^{ijk}A^j_\mu A^k_\nu$ (see Equations (8.13a) and (8.13b) in [21]). The same reduction can be carried out for pure Yang–Mills SU(3) fields. Clearly, the kinetic (i.e., pure gauge) terms in TQFTs can be reduced from the exchange terms in the corresponding lattice gauge theories. The TQFTs inherit the basic characters of lattice gauge theories, such as the noncommutative operators (of Clifford algebras), nonlocality, and nontrivial topological structures (caused by dimensionality and many-body interactions), which also require the time average with monodromy representation, violating the ergodic hypothesis.

According to Corollary 1, the TQSM must take into account the contribution of nontrivial topological structures to physical properties of the system by performing a gauge transformation in higher dimensions. The nontrivial topological structures observed in the Hamiltonian and the partition function of TQSM models (such as the 3D Ising model, the 3D $Z_2$ lattice gauge theory, etc.) can be found to be hidden in the Lagrangian and the partition function of corresponding TQFTs. Performing the integrals of these field theories actually involves the interchange of many-body interacting particles, which should account for the contribution of nontrivial topological structures and generate the topological phases in the systems for calculations of physical properties. From the facts above, and as the immediate consequence of Corollary 1, the following corollary is validated:

**Corollary 5.** *The topological quantum field theories must take into account the contribution of nontrivial topological structures by performing a gauge transformation or a monoidal transformation in one-dimensional higher space, while topological phases are generated in field operators [11–13].*

According to Definition 3, TQFTs are specific models in QFTs, in which the nontrivial topological structure and the long-range spin entanglement exist, which should be set up within the framework of QM. The JNW theorem has been applied to various QM theories [36–39] and QFTs [114–117]. The logic flow can be represented as follows: QM ⊃ QFT ⊃ TQFT and, because JNW for QM, we have JNW for QM → JNW for QFT → JNW for TQFT. Clearly, the JNW framework as the mathematical basis of quantum mechanics provides the sufficient condition that it is the mathematical basis of topological quantum field theories.

Then, we show that it is necessary to use the JNW framework for TQFTs. We first take a $\varphi^4$ scalar field theory as an example. As mentioned above, the kinetic (i.e., pure gauge) terms in the $\varphi^4$ quantum field theory in (3+1)D space–time with ultraviolet cutoff can be mapped to the 3D Ising model (and the 3D $Z_2$ lattice gauge theory) [21,110]. The lattice theories with ultraviolet cutoff can approach the continuous limit with the lattice space a → 0. The potential terms (for instance, $\varphi^2$ and $\varphi^4$ in the $\varphi^4$ scalar field theory) can be treated as Higgs-like terms [63,64,68–72,77,78]. The kinetic (i.e., pure gauge) terms possess the anti-communicative relations between spinor operators, which require the use of Jordan algebra within the JNW framework. In the previous work [11], we have already shown that the 3D Ising model must be dealt with in the (3+1)D JNW framework. Therefore, a $\varphi^4$ quantum field theory in (3+1)D space–time must be investigated in a (3+2)-dimensional space–time. For the (3+1)D $\varphi^4$ quantum field theory, an additional time dimension is added, together with the imaginary time, to form the complex time, in order to fit the JNW framework. Similarly, the (3+1)D Yang–Mills quantum field theory with gauge symmetry SU(2) or SU(3) [2–6,92–98] must be studied also in the (3+2)D space–time. When we study TQFTs, no matter whether we need to consider the gauge terms with/without Higgs(-like) terms, we have to deal with noncommutative relations (Γ-matrices of Clifford algebra, non-Abelian of spin operators), which can be solved by Jordan algebra. The commutativity of field operators with Jordan algebra and generalized Yang–Baxter equation together ensure the integrability of TQFT models [11,16,67]. According to Corollary 2, the TQSM must be set up on the JNW framework. From the observations above, and as the immediate result of Corollary 2, the following corollary is held:

**Corollary 6.** *The topological quantum field theories must be set up on the Jordan–von Neumann–Wigner framework, with application of Jordan algebras for multiplication of field operators, which ensures the integrability of the system.*

In brief, according to Corollary 3, the ergodic hypothesis is violated at the finite temperature in TQSM. In TQFTs with kinetic terms (i.e., pure gauge terms) mapped from the models in TQSM, the ergodic hypothesis is violated also at the finite temperature. This conclusion is validated for TQFTs with/without Higgs(-like) terms. Thus, we have demonstrated the following corollary:

**Corollary 7.** *The ergodic hypothesis is violated at the finite temperature in the topological quantum field theories.*

**Remark 5.** *Corollaries 6 and 7 are complementary in a sense that the JNW framework provides the mathematical basis of TQFTs, while it actually performs a time average for a physical quantity, violating the ergodic hypothesis.*

According to Corollary 4, the time average of the ensemble average and the quantum mechanical average of any physical quantity f equals the value obtained by an appropriate measurement on the given system. Besides the mapping $\beta = (k_B T)^{-1} \to it = \tau$ by performing the Wick rotation, the temperature–time duality has a new insight, with the symmetric formulations for the time average and the temperature average [97]. According to our previous results [10–13], the time averaging must be used in the 3D Ising model to address it in (3+1) dimensions, and a temperature–time duality exists. The time averaging of a function $\varphi(x(t, \tau))$ is obtained by evaluating the integral:

$$\langle\phi(x(t,\tau))\rangle_t = \frac{1}{\Delta t} \int_{t_0}^{t_0+\Delta t} dt\phi(x(t,\tau)) \qquad (13)$$

where τ represents a real variable going from 0 to β. The experiment can be performed over a period of time $t_0 < t < t_0 + 9\Delta t$. In the systems like the 3D Ising model, the contributions of the nontrivial topological structures to the partition function (and the correlation functions and other physical quantities) cannot be neglected. The time necessary for the time averaging must be infinite, which is comparable with or even much larger than the time of measurement of the physical quantity of interest. Therefore, the ergodic hypothesis may be invalid, $\Delta t \to \infty$. With the mapping of t ↔ -iτ (or τ ↔ it) and a transformation of variances (-iτ ↔ τ' and it ↔ t'), the time averaging is equivalent to the temperature averaging, which can be written as [112]:

$$\langle\phi(x(\tau',t'))\rangle_{\tau'} = \frac{1}{\Delta\tau'} \int_{\tau'_0}^{\tau'_0+\Delta\tau'} d\tau'\phi(x(\tau',t')) \qquad (14)$$

In Equation (16), the imaginary unit i does not explicitly appear because it is rescaled by the variance transformation. It is worth noting that the temperature averaging is performed somehow in the imaginary time-parametric space. The temperature–time duality is clearly seen from Equations (15) and (16). However, we should remember that, besides the temperature average for τ = it, we also have to account for the time average for t (according to Corollaries 3 and 4). In TQSM, the 3D $Z_2$ lattice gauge theory is a dual model of the 3D Ising model, which must be dealt with in the parameter space of complex time (t + it) (or complex temperature (τ–iτ)). Other Abelian U(1) and non-Abelian SU(2) or SU(3) lattice gauge theories are the generalizations of the 3D $Z_2$ lattice gauge theory, which, thus, should follow this role. Because the actions of these lattice gauge theories can be mapped to the kinetic (i.e., pure gauge) terms in TQFTs, the parameter space of complex time (or complex temperature) must be used also in these field theories. This means that the (3+1)D TQFTs must be investigated in (3+2)D dimensions.

Therefore, in consideration of both the temporal integral in the imaginary time of the Lagrangian of TQFTs (including the $\varphi^4$ scalar field theory, the Abelian gauge field theory, the Yang–Mills SU(2) or SU(3) gauge field theory) and the addition of the time dimension for time average to deal with the nontrivial topological structure, the TQFTs must deal with the parameter space of complex time (t, τ), with the twofold temporal integrals of t and τ. It is clear that, for TQFTs, the complex time averaging of a function φ(x(t, τ)) is expressed by evaluating the integrals:

$$\langle\langle\phi(x(t,\tau))\rangle_\tau\rangle_t = \langle\langle\phi(x(t,\tau))\rangle_t\rangle_\tau = \frac{1}{\Delta t}\int_{t_0}^{t_0+\Delta t}\frac{1}{\Delta \tau}\int_{\tau_0}^{\tau_0+\Delta \tau} dtd\tau\phi(x(t,\tau)) \quad (15)$$

Equation (17) is validated for the functions satisfying Fubini's theorem and it is true for the Lagrangian of TQFTs. According to the temperature–time duality, the parameter space of the complex time (t, τ) can be dual to the parameter space of complex temperature (τ, t). In the $\varphi^4$ quantum field theory and also other TQFTs, the real-time average of the temperature (or imaginary time) average of a function φ(x(t, τ)) is identical to the temperature (or imaginary time) average of the real-time average of the function φ(x(τ,t)).

With the preliminaries above, we can claim the following corollary:

**Corollary 8.** *The topological quantum field theories must deal with the parameter space of complex time (or complex temperature). The real-time average of the temperature (or imaginary time) average of a function φ(x(t, τ)) is identical to the*

*temperature (or imaginary time) average of the real-time average of the function φ(x(t, τ)).*

Finally, we prove the following theorem:

**Theorem 4**. *In models in the topological quantum field theories, a topological phase transition occurs near the infinite temperature, which shows a symmetrical breaking of time inverse symmetry. It corresponds to the change in the parameter space from imaginary time (real temperature) to complex time (or complex temperature).*

**Proof of Theorem 4.** The kinetic (pure gauge) terms in QFTs show the nontrivial topological structure at the finite temperature due to the interactions between spins in 3D (according to our results in the 3D Ising model [10–13]). Thus, it is easy to accept that a topological phase transition occurs near the infinite temperature in these field theories, which shows a symmetrical breaking of time inverse symmetry. Such a topological phase transition is similar to what happens in models of TQSM, as proven for Theorem 3. The energies needed to generate these new bosons are much higher than what we need in order to generate the massive Higgs bosons [63,64,68–72,77,78], since these massless bosons emerge near the infinite temperature, whereas the Higgs bosons generate at the critical point $T_c$ with Higgs terms in the Lagrangian or Hamiltonian. Near the infinite temperature, the parameter space from imaginary time (real temperature) is changed to complex time (or complex temperature) with decreasing temperature and, as a result, follows Corollary 8. □

Obviously, we have the following remark:

**Remark 6**. *The mathematical structures of TQFTs are realized also as a result of interplay between algebra, topology, and geometry.*

The topological/geometrical phases can be generated during inter-exchanging the particles in spacetime in TQFTs, which are closely connected to the Jones polynomial [27,118] with the formulas of Wilson loop [119,120] and Witten integral [95–98] for the action of the gauge group [2–6]. These phases are analogous to those in the Aharonov–Bohm effect [121,122], the Berry phase effect [92,105,123], fractional quantum Hall effect [123,124], etc., with the mathematical basis of Chern's geometric theory [122,125,126]. A detailed discussion for the mathematical structures of the 3D Ising model can be found in [13]. It is worth highlighting that twofold integrations on the curvature of the 3D Ising model in spacetime result in the Gauss–Bonnet–Chern formula for the Chern number $c_1(F)$ of the first class, from which one can fix some topological invariants, such as the winding numbers and the topological phases [13]. According to Corollary 8, twofold integrations in parametric space of complex time (or complex temperature) on the curvature of other TQFT models will also result in topological invariants in connection with the Chern numbers, the winding numbers, and the topological phases.

It is worth noting that Zeng [113] proposed a naive but natural idea that time may emerge as the holographic dimension of gauge systems in Euclidean space, which takes a statistic model, e.g., the 3D Ising model [10], as concrete implementations. Based on the so-called statistic/gravity duality, Zeng [113] proposed the idea that the time flow of (n + 1)D universe is to identify the renormalization group flow (inverse if necessary) of some nD statistic models at equilibriums. Thus, it makes time emerge naturally from lower dimensional physic systems without time concepts. For 3D Ising model, the dual universe with emergent time flow has qualitatively the same evolution history as the real world. The idea of identifying the renormalization group flow with the cosmic time flow projects remarkable highlights on the solving of the cosmological constant problem. If this idea is the case, then the acceleration of (3+1)D universe has relevance only with properties of some 3D equilibrium statistic models or Euclidean field theories, while the vacuum energy of (3+1)D quantum field theories could affect dynamics only of some (3+2) or (4+1)D universes. This is consistent with our results for an emergent time axis in the 3D Ising model [10,11,16,17,20,59], and also in TQSM and TQFTs.

**5. Conclusions**

In conclusion, we build the framework of TQSM and TQFTs with respect to the mathematical aspects (topology, algebra, and geometry) and physical features (the contribution of topology to physics, JNW framework, time average, ensemble average, and quantum mechanical average) by a generalization of our findings and observations in the 3D Ising models. We found that, in these theories, the nontrivial topological structures exist, which contribute to thermodynamic physical properties of the many-body interacting spin (or particle) systems. It is found also that these theories must be set up within the JNW framework, and the ergodic hypothesis is violated at the finite temperature in these theories. It is necessary to account for the time average of the ensemble average and the quantum mechanical average in TQSM and to introduce the parameter space of complex time (and complex temperature) in the TQFTs. Furthermore, we uncover that, in models in these theories studied in this work, a topological phase transition occurs near the infinite temperature, with a symmetrical breaking of time inverse symmetry and the emerging of massless gauge bosons. We also showed that the mathematical structures of TQSM and TQFTs are realized as a result of interplay between algebra, topology, and geometry.

**Funding:** This research was funded by the National Natural Science Foundation of China under grant number 52031014.

**Acknowledgments:** The author is grateful to Fei Yang for understanding, encouragement, support, and discussion.

**Conflicts of Interest:** The authors declare no conflict of interest.